\documentclass[a4paper]{aa}
\usepackage{mybib, epsfig, times}

\begin{document}
\thesaurus{3(12.03.1; 11.09.3; 11.17.1; 11.17.4 Q0000$-$2619)}

\title{Constraints on the physical properties of the 
	damped Ly$\alpha$ system of Q0000$-$2619 at $z$ = 3.054}

\author{G.~Giardino \and F.~Favata}

\institute{Astrophysics Division -- Space Science Department of ESA, ESTEC,
  Postbus 299, NL-2200 AG Noordwijk, The Netherlands}

\offprints{G. Giardino} \mail{ggiardin@astro.estec.esa.nl}

\date{Received date; accepted date}

\titlerunning{Constraints on the physical properties...}
\authorrunning{G.Giardino \& F. Favata}

\maketitle 
\begin{abstract}
  
We present the detection of C\,{\sc ii} and  C\,{\sc ii}$^*$
absorption in the $z = 3.0543$ damped Ly$\alpha$ system toward Q0000$-$2619.
The derived population ratio implies a fine structure excitation
temperature between 19.6 and 21.6 K. The upper value sets a strict upper limit on
the CMB temperature at this redshift, which is consistent with the
predicted value of 11.05 K from standard cosmology.
Under the assumptions of an ionization degree
ranging from 0 to 10\%, a gas kinetic temperature between 100
and 10\,000 K and a UV field with a Milky Way spectrum,  
the density of the absorber is constrained to be between 0.7 and 40
cm$^{-3}$ and the H-ionizing flux between 1 and 80 times the intensity
of the Galactic UV field. If the damped Ly$\alpha$ system is assumed to be
homogeneous, the implication is that its size in the direction of the line of sight
must be between 1 and 100 pc.  


\end{abstract}

\section{Introduction}
\label{sec:intro}

\begin{figure}[htbp]
  \begin{center}
    \leavevmode \epsfig{file=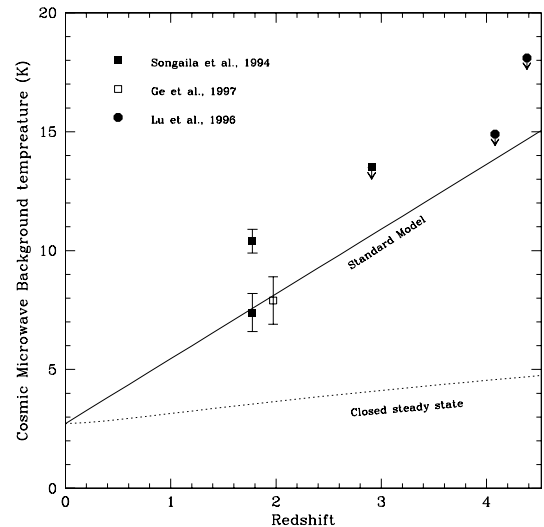, width=8.5cm, bbllx=10pt,
      bblly=150pt, bburx=600pt, bbury=700pt, clip=}
    \caption{Upper limits to the CMB temperature at $z>0$. The
predicted relations between the CMB temperature and the
redshift, for the Standard Model (the hot Big Bang) and for the Closed
Steady State Model (Phillips\,1994) are also shown.}
    \label{fig:TcmbULall}
  \end{center}
\end{figure}

Fine structure transitions observed 
in the absorption spectra of quasars
provide unique information on the temperature of the microwave background
at the redshift of the absorber, on the intensity of the UV-field and
on the density of the absorbing system (\cite{Bahcall68}).  
The measured excitation temperatures, or upper limits to the excitation
temperature, of the fine structure of 
C\,{\sc i} and C\,{\sc ii} has been used by several authors to
constrain the temperature of the Cosmic Microwave Background radiation
(CMB) up to
a redshift of 4.38 (\cite{Lu96}). The most significant published constraints on
the  CMB temperature at $z > 0$ are
summarized in Fig.\,\ref{fig:TcmbULall}. The Big Bang cosmological model predicts a
simple relationship between the CMB temperature and the redshift $z$
(e.g. \cite{Peebles93}): 
\begin{equation}
T_{\rm CMB}(z) = T_{\rm CMB}(0)(1+z)
\end{equation}
Alternative anisotropic cosmological models (\cite{Phillips94}) 
make strong claims for a value of $T_{\rm CMB}$ more than 5 K below the
standard prediction at $z = 3$.
So far all the measured
excitation temperatures and upper limits are consistent with a Friedmann universe.

The observed excitation temperature (or upper limit) of C\,{\sc i} and
C\,{\sc ii} fine structure has also been used to constrain the density
of the absorbing systems (\cite{Chaffee88}; \cite{Songaila94}; \cite{Ge97}). 

Here we discuss the detection of C\,{\sc ii} and  C\,{\sc ii}$^*$
absorption at 
$z =3.0543$ in the spectrum of Q0000$-$2619 obtained
at the NTT. The absorption lines are associated with a damped Ly$\alpha$
system with neutral hydrogen column density of 
$N$(H\,{\sc i})$ = 1.5 \pm 0.5 \times 10^{20}~{\rm cm^{-2}}$
(\cite{Savaglio94}, SOM hereafter).
By measuring the equivalent width of the C\,{\sc ii} multiplet absorption 
lines we derive strict upper limits on the temperature of the CMB at this
redshift and constrain the density of the absorbing systems.
In the next section the observations are briefly presented. In Sect.\,\ref{sec:Texc} the 
C\,{\sc ii} fine structure excitation temperature, which  gives directly the upper
limit on the CMB temperature at redshift 3.0543, is derived.  In
Sect.\,\ref{sec:mechanisms} the relative strengths of the different
excitation mechanisms are reviewed.
The constraints on the density and the UV field
in the  $z = 3.0543$ damped Ly$\alpha$ system are derived in
Sect.\,\ref{sec:results} and discussed in Sect.\,\ref{sec:disc}. 

\section{Observations}
\label{sec:obs}

In October 1990 echelle observations of Q0000$-$2619
at $z = 4.12$ were obtained with the ESO Multi Mode Instrument (EMMI)
(\cite{Dodorico90}) at the ESO NTT telescope. The spectra cover the
wavelength range from 4400 \AA\ to 9265 \AA\  with a resolution of 0.2
and 0.3
\AA\ between 4700 and 8450 \AA\ and signal-to-noise ratio S/N$=15-60$ per resolution
element.
The data were reduced
and analysed by \cite*{Savaglio97} to which we refer the reader for a
detailed description of the observations and data reduction procedure.
In the spectrum of Q0000$-$2619, SOM have identified nine metal absorption systems; among
these two are known damped systems at redshifts 3.054 and 3.390. 
Eight of the nine systems have redshift greater than 3.
We carefully inspected the spectrum looking for absorption from the 
C\,{\sc i} and  C\,{\sc ii} ground state multiplet. For all systems at
redshifts greater than 3,  the absorption from
the C\,{\sc i} ground state multiplet would
land redwards of the Lyman-$\alpha$ emission.  No absorption from
C\,{\sc i} is detectable in the spectrum. 

C\,{\sc ii} absorption was detected for both the
damped systems at $z=3.054$ and $z=3.390$ (SOM). 
For these systems the C\,{\sc ii}~($\lambda1334$) absorption line lands
bluewards of the Ly$\alpha$ emission.
The  C\,{\sc ii} at $z =  3.3913$ is heavily blended, while the C\,{\sc ii} at
redshift $z = 3.0543$ is reasonably clean, despite falling in the Lyman forest at $\lambda = 5410.6$. 
At $\lambda = 5415.4$ we detect a weak  
absorption line (3.5$\sigma$  confidence) 
consistent with absorption from the excited
fine-structure level of C\,{\sc ii} at
redshift $z = 3.0543$. 

Table \ref{tab:lines} summarizes the data on the absorption lines
detected for the system at $z =3.0543$. 

\begin{table}
\begin{center}
\footnotesize
\begin{tabular}{c c c c c c}
\hline
FWHM$^1$ 		& $\lambda_{\rm obs}$	& $W_{\lambda}$	& $\sigma(W_{\lambda})$	& $z_{\rm abs}$	& ID 	\\ 
({\rm km/s})	& (\AA)			& (\AA)	& (\AA)		&		&	\\
\hline
11		&$\sim$4928		& $>30$	& ...		&$\sim$3.054	& Ly$\alpha$ \\
11		&5279.65		&1.79	& 0.17		&3.0545		& OI(1302)\\
11		&5410.54		&1.16	& 0.02		&3.0543		& CII(1334)\\
11		&5415.37		&0.08	& 0.02		&3.0543		& CII$^*$(1335)\\
11		&5650.72		&0.57	& 0.09		&3.0544		& SiIV(1393)\\
11		&5687.20		&0.15	& 0.08		&3.0543		& SiIV(1402)\\
11		&6276.49		&0.06	&0.02		&3.0541		& CIV(1548)\\
11		&6287.08		&0.03	&0.02		&3.0542		& CIV(1550)\\
\hline
\end{tabular}
\end{center}
$^1$ Spectral resolution
\caption{The damped Ly$\alpha$ system at $z_{\rm abs} = 3.0543$.  CII and CII$^*$ wavelengths
and equivalent
widths are from this paper, all the other line measurements and
identifications are from SOM.}
\label{tab:lines}
\end{table}

\begin{figure}[htbp]
  \begin{center}
    \leavevmode \epsfig{file=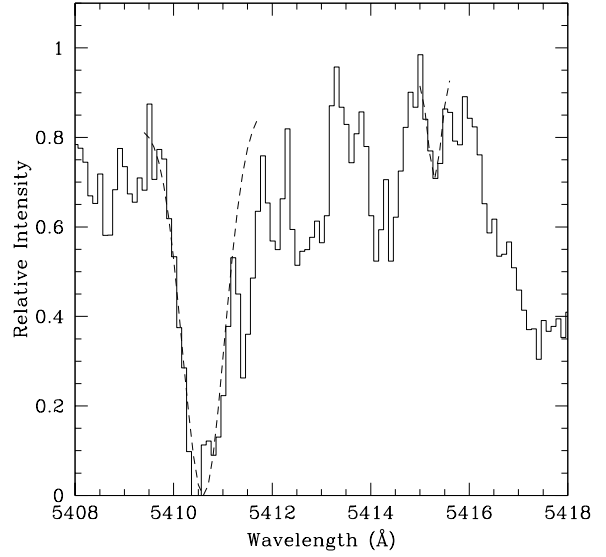, width=8.0cm, bbllx=10pt,
      bblly=150pt, bburx=600pt, bbury=700pt, clip=}
    \caption{The spectrum of Q0000$-$2619 in the 
vicinity of the C\,{\sc ii} multiplet at $z = 3.0543$ with profile fit to
    the C\,{\sc ii}~$\lambda1334.53$ and  C\,{\sc
    ii}$^*$~$\lambda\lambda1335.66,1335.71$ lines. The dashed line is the Gaussian fit
    to the absorption lines.}
    \label{fig:Cii}
  \end{center}
\end{figure}

\section{Excitation temperature}
\label{sec:Texc}

At a redshift of 3.0543, the C\,{\sc
ii} $J = 1/2$ ($\lambda$1334.53 \AA) and  $J = 3/2$
($\lambda\lambda$1335.66,1335.71 \AA) absorption lines land respectively
at 5410.54 \AA\ and 5415.17,5415.37 \AA.
Fig.\,\ref{fig:Cii} shows the spectrum of Q0000$-$2619 in the
vicinity of the C\,{\sc ii} multiplet. In this wavelength range the spectral resolution
is of 0.2 \AA\ and the S/N per resolution element is about 15.	

\begin{figure}[htbp]
  \begin{center}
    \leavevmode \epsfig{file=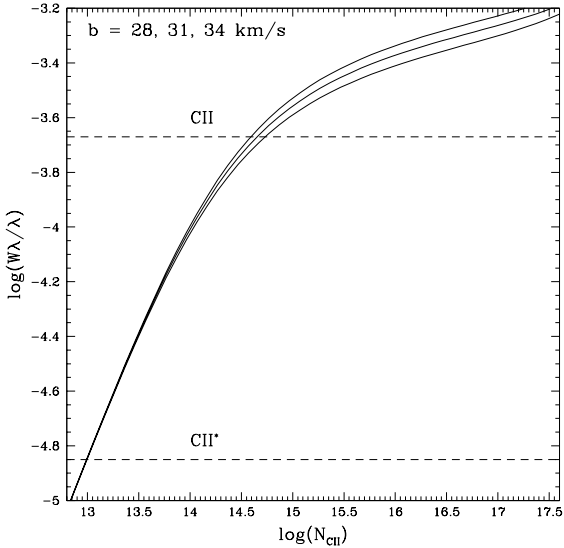, width=8.0cm, bbllx=10pt,
      bblly=150pt, bburx=600pt, bbury=700pt, clip=}
    \caption{Theoretical curve of growth for C\,{\sc ii}, for $b =$
    28, 31 and 34 km/s. The dashed lines are the value of
    $\log(W_\lambda/\lambda)$ measured for  C\,{\sc ii} and C\,{\sc
    ii}$^*$. C\,{\sc ii} $J = 1/2$ ($\lambda$1334.53 \AA) and  $J = 3/2$
	($\lambda\lambda$1335.66,1335.71 \AA) absorption lines 
	have oscillator strength 0.128, 0.013 and 0.115 respectively (Morton~1991).}
    \label{fig:cg}
  \end{center}
\end{figure}

The multiplet
lands in the Lyman forest and  in the damping wing of the
Ly$\alpha$ absorption of the $z =  3.390$ 
damped system, at $\sim$ 5337 \AA.
The C\,{\sc ii} ground state absorption line is
slightly blended. 
We used a multiple Gaussian fit to deblend the C\,{\sc ii} line
and measured  an equivalent width of $W_{\lambda} = 1.16 \pm 0.02$
\AA. The C\,{\sc ii}$^*$ absorption line is detected
at $3.5\sigma$. For C\,{\sc ii}$^*$ $W_{\lambda} = 0.077 \pm 0.02$. 
The Gaussian fit to the line has a FWHM of 0.29 \AA\ which is consistent
with the instrument resolution.
The two  equivalent widths correspond  respectively to 
$\log(W_\lambda/\lambda) =  -3.67$ and $-4.85$. In both cases a local
continuum level corresponding to the damping wing of the
Ly$\alpha$ absorption at $z =  3.390$ has been used.
The C\,{\sc ii}
absorption line is well fitted by a Gaussian having a FWHM of $53\pm5$ km s$^{-1}$,
which given the instrumental resolution in this range of 11 km~s$^{-1}$
corresponds to an intrinsic $b$ parameter of $31 \pm 3$ km s$^{-1}$.
In Fig.\,\ref{fig:cg}  the theoretical curve of growth of C\,{\sc ii} is
plotted for three values of $b$: 28, 31 and 34 km s$^{-1}$. The two
values of $\log(W_\lambda/\lambda)$ are also shown.
The corresponding column density values
derived for  C\,{\sc ii}  and C\,{\sc ii}$^*$ are summarized in Table \ref{tab:CII}. 

\begin{table}
\begin{center}
\footnotesize
\begin{tabular}{c c c c }
\hline
$b$(km/s)				& 28		& 31		& 34 			\\ 
\hline
$N$(C\,{\sc ii})($\times 10^{14}{\rm ~cm^{-2}}$)	& $5.0\pm0.3$ 	& $4.3\pm0.3$	& $3.5\pm0.2$\\  
$N$(C\,{\sc ii}$^*$)($\times 10^{13}{\rm ~cm^{-2}}$)  	& $1.0\pm0.2$	& $1.0\pm0.2$	& $1.0\pm0.2$\\
\hline							
$T_{\rm ex}(K)$				& $19.6\pm1.1$	& $20.5\pm1.2$ 	& $21.6\pm1.2$\\
\hline
\end{tabular}
\caption{C\,{\sc ii} fine structure population and excitation temperature.}
\label{tab:CII}
\end{center}
\end{table}

According to the Boltzmann equation, an excitation temperature $T_{\rm ex}$
can be expressed in terms of the column densities $N_1$ and $N_0$ in
the excited and the ground-state level:
\begin{equation}
N_1/N_0 = g_1/g_0 \exp(-\Delta T_{10}/T_{\rm ex})
\label{eq:boltz}	
\end{equation}
where $k\Delta T_{10}$ is the energy difference between the excited
level (1)  and the ground level (0). 
For the fine structure levels $J= 3/2$ and $J=1/2$ of C\,{\sc ii},
$\Delta T_{10} = 91.2$ K. The weights $g_J$ are given by $2J + 1$.
The derived ratio of column
densities, $N_1(J=3/2)/N_0(J=1/2)$ corresponds to
an excitation temperature of 19.6 K (for $b=28$ km s$^{-1}$),  20.5 K
(for $b=31$ km s$^{-1}$) and
21.6 K (for $b=34$  km s$^{-1}$).
The excitation temperature of 21.6 K provides
a strict upper limit on the temperature of the CMB at the absorber
redshift of 3.0543; 
this upper limit would hold even if the C\,{\sc
ii}$^*$ absorption were a spurious effect of the Lyman forest.
The CMB temperature at this redshift 
is predicted to be 11.05 K by the Big Bang model.

The excitation temperature provides a strict upper limit to the CMB
temperature as other sources may contribute appreciably to the
excitation. In fact, if the C\,{\sc ii}$^*$ absorption is not a spurious
effect of the line forest, other mechanisms have to be at play to
explain an excitation temperature of at least 19.6 K.
The possible excitation mechanisms are reviewed in the next
section.

\nocite{Morton91}

\section{Excitation mechanisms}
\label{sec:mechanisms}

The higher fine-structure states of a
ground state multiplet can be populated by (1) particle collisions,
(2) direct excitation by infrared photons and (3)
indirect excitation by ultraviolet photons. 

The equilibrium between the excitation and de-excitation of the
C\,{\sc ii} $ J = 1/2 \rightarrow 3/2$ fine structure is 
described by:

\begin{eqnarray}
N_0(\sum_j \langle\sigma_{01}v\rangle n_j + B_{01}U(\nu_{01}) +
\Gamma_{01}) = \\
~~~~~~N_1(A_{10} + \sum_j \langle\sigma_{10}v\rangle n_j+
B_{10}U(\nu_{01}) + \Gamma_{10}) \nonumber
\label{eq:balance1}
\end{eqnarray}
Here the collision excitation rate is expressed as
$\langle\sigma v\rangle n$, where $\langle\sigma v\rangle $ is
the temperature averaged product between the 
cross-section and the particle velocity and $n$ is the particle
density, $j=$ H, $e,~p$.  The direct photon excitation rate  is expressed as the
product between the Einstein probability coefficient for induced transition 
($B$) and the radiation energy density per frequency interval
($U(\nu_{01})$). The UV  pumping rate  is represented
by the coefficient $\Gamma$, which includes the UV energy density term. $A_{10}$ is the 
Einstein probability coefficient for spontaneous transition and 
for the C\,{\sc ii} transition $J = 3/2 \rightarrow 1/2$, $A_{10}
= 2.29\times 10^{-6}$s$^{-1}$ (\cite{Nussbaumer81}). For convenience
we can divide both sides of Eq.\,(3) by $A_{10}$ and rewrite:
\begin{eqnarray}
N_0(\sum_j q_{01,j}n_j + b_{01}U(\nu_{01}) + \gamma_{01}) = \\ 
~~~~~~N_1(1 + \sum_j q_{10,j}n_j + b_{10}U(\nu_{01}) + \gamma_{10}) \nonumber
\label{eq:balanceN}
\end{eqnarray}
where $q_{ij,j}$, $b_{ij}$ and $\gamma_{ij}$ are respectively
the collisional excitation rate, the Einstein
probability coefficient for induced transition and the UV pumping
rate, all divided by $A_{10}$.
In order to establish the importance of the various factors in determining the
observed population ratio in the fine structure levels of C\,{\sc ii}, 
we need to evaluate the magnitude of each term in this equation.

\subsection{Collisional excitation and de--excitation}

The particles which may be responsible for
collisional excitation are essentially electrons, protons and atomic
hydrogen, with different contributions dominating in different
temperature and ionization regimes. 

The expression for the rate of collisional
excitation by electrons is given in \cite*{Bahcall68} and, by
using the detailed computation of the effective collisional strength
given in \cite*{Hayes84} and \cite*{Keenan86}, one  
obtains typically $q_{01, e^-} = 0.15$ ${\rm cm^3~s^{-1}}$ for $T_e =
100$ K and $q_{01, e^-} = 0.051$  ${\rm cm^3~s^{-1}}$ for $T_e = 10\,000$ K.

The excitation rates as a function of electron temperature for neutral
hydrogen collisions are also given in \cite*{Keenan86} and for $T_e =
100$ K
$q_{01, {\rm H}} = 2.8 \times 10^{-4}$ ${\rm cm^3~s^{-1}}$, 
for $T_e = 10\,000$ K $q_{01,{\rm H}} = 1.5 \times 10^{-3}$ ${\rm cm^3~s^{-1}}$.
The electron collision term will dominate the hydrogen
collision term for $n_e \geq 0.002~n_{\rm H}$  at $T_e =
100$ K and for $n_e \geq 0.03~n_{\rm H}$ at $T_e = 10\,000$ K. 
If the plasma is collisionally ionized the electron collision term will be the dominant one for
temperature $T > 10\,000$ K, when the fraction of ionized hydrogen
becomes significant (\cite{Bahcall68}).  If the absorbing medium
is significantly photoionized and $n_e \geq 0.03~n_{\rm H}$ 
the electron collision term will be the dominant
term at any kinetic temperature whereas if
photoionization is not signficant and $n_e <
0.002~n_{\rm H}$ hydrogen collisions will dominate.

The C\,{\sc ii} excitation rate for proton collision becomes
comparable to the electron contribution only for temperatures $T_e \geq
10^5$ K (\cite{Bahcall68}),  but at these
temperatures C\,{\sc ii} is completely destroyed by collisional
ionization (\cite{Sutherland93}). 
We will therefore ignore the proton collision contribution.  

The collisional de-excitation rate is given by:
\begin{equation}
\langle\sigma_{10, j}v\rangle = \frac{1}{2} 
\langle\sigma_{01,j}v\rangle \exp(91.2 ~{\rm K}/T_e)
\label{eq:sigma_dex1}
\end{equation}
As a first approximation, the collisional de-excitation 
term on the right hand side of Eq.\,(4)
can be omitted if $n_e < 1$ cm$^{-3} $ or 
$n_{\rm H} < 10^3$  cm$^{-3} $.

\subsection{Direct IR photon excitation and de--excitation}

The photons responsible for directly populating C\,{\sc ii} fine structure
excited levels have a frequency of $\bar{\nu}_{10} = 64.0$ cm$^{-1}$. 
Sources of far--infrared photons are the CMB and thermal dust emission. 
Since the CMB radiation has  a Black Body spectrum the 
direct excitation rate from CMB photons can be expressed as:
\begin{equation}
b_{01}U(\nu_{01}) = 2 \exp(-91.2~{\rm K} / {T_{\rm CMB}})
\label{eq:sigma_dex2}
\end{equation}
As we have seen, the
Big Bang model predicts the CMB temperature to be 11.05 K at this
redshift, moreover the upper limit on the CMB temperature  at 
$z = 4.08$ set by \cite*{Lu96}, constrains the CMB temperature empirically. 
If we assume that the CMB temperature varies monotonically with $z$:
\begin{equation}
T_{\rm CMB}(z) = T_{\rm CMB}(0)(1+z)^\alpha
\end{equation}
then the measure of \cite*{Lu96} gives $\alpha < 1.05$ between $z =0$
and $z = 4.08$, that is $T_{\rm CMB} < 11.85$ K at $z = 3.0543$.
This implies that $b_{01}U(\nu_{01})$ must be lower than $ 9.1 \times
10^{-4}$~s$^{-1}$, i.e.  only a minor contribution to the derived
population ratio in our system, where $N_1/N_0$ for C\,{\sc ii} is
between 0.02 and 0.03 (Table \ref{tab:CII}). As a first approximation
therefore this term can be neglected and
the same is true for $b_{10}U(\nu_{01}) = \frac{1}{2} b_{01}U(\nu_{01})$.

An infrared photon flux with an intensity comparable to the one measured
in the Galactic plane (\cite{Bennett92}, \cite{Kogut96}) 
would correspond to an excitation rate at least
2 orders of magnitude smaller than the one for a CMB photon flux with $T_{\rm CMB}
= 10$ K and can be ignored.

\subsection{Indirect UV photon excitation and de--excitation}

The other important type of photon excitation is UV photon
pumping. After the absorption of a photon an atom will usually cascade back
through a variety of states, sometimes reaching levels that could not be
populated by direct radiative upward transition from the ground
state.
If $m$ represents all the quantum numbers for one of the upper
levels, reached by photon absorption, the transition rate from level
0 to level 1, is given by (\cite{spitzer}):
\begin{equation}
\Gamma_{01} = \sum_m  B_{0m}U(\nu_{0m})\epsilon_{m1}
\label{eq:UVpumping1}
\end{equation}
where $\epsilon_{m1}$ is the fraction of downward transitions from
level $m$ that populate level 1, when the atom first reaches the group
of lower levels. For transitions within a multiplet the values of
$\epsilon_{mj}$ are tabulated (e.g. \cite{allen}).
To evaluate $\Gamma_{01}$ we considered all the direct upward transitions
from C\,{\sc ii} ground state $2p^2P^0$, longwards  of 900 \AA: 
$^2P^0 \rightarrow$ $^2P$ ($\lambda\lambda903,904$), $^2P^0 \rightarrow$ $^2S$ ($\lambda\lambda1037,1036$),
$^2P^0 \rightarrow$ $^2D$ ($\lambda\lambda1335,1334$).
For the UV field we adopted the Milky Way spectral energy distribution
(SED) given by \cite*{Black87} with a Milky Way intensity at 912 \AA\ of 
$4.7 \times 10^{-19}{\rm erg~cm^{-2}~s^{-1}~Hz^{-1}}$ (\cite{Mathis83}),
obtaining:
\begin{equation}
\Gamma_{01} = 5.3 \times 10^{-10}~{\rm s^{-1}},
\label{eq:UVpumping2}
\end{equation}
that is $\gamma_{01} = 2.3\times 10^{-4}$~\footnote{\cite*{Keenan86} 
obtained $\Gamma_{01} = 2.4 \times 10^{-10} {\rm
s^{-1}}$ by using the UV intensity field and SED given in
\cite*{Gondhalekar80}}. 
Since the UV pumping de-excitation rate $\gamma_{10}$ 
is of the same order of magnitude as $\gamma_{01}$, 
the UV de-excitation term  can also be omitted in
Eq.\,(4), for any likely UV flux intensity and SED.

The final shape of the balance equation is:
\begin{equation}
N_1/N_0 = \sum_j q_{01,j} n_j + \gamma_{01}
\label{eq:finalb}
\end{equation}
with the collisional term being dominated by the electrons or hydrogen
atoms contribution according to the absorber ionization degree. 
The UV field can also be expressed in terms of the hydrogen density
through the ionization parameter, $U = \phi({\rm H})/n_{\rm H}c$, where $\phi({\rm H})$
is the surface flux of hydrogen-ionizing photons. If we assume a UV
flux having the Milky Way SED given in \cite*{Black87},
$ \gamma_{01} = 0.7 n_{\rm H} U$ and:
\begin{equation}
T_{\rm ex} = \frac{-91.2~{\rm K}}{\log[0.5(\sum_j q_{01,j} f_j n_{\rm H}
+ 0.7 n_{\rm H} U)]}
\label{eq:final}
\end{equation}
where $f_j$ is the fraction of particle $j$ with respect to the hydrogen
density. 

\section{Results} 
\label{sec:results}

From the ratio C\,{\sc ii}/C\,{\sc iv} observed in our data at $z=3.0543$ SOM 
derived  $\log U \geq -3.2$ and assuming solar abundance ratios a
consistent fit to the data was obtained with $\log U = -2.8$ and 
$Z \sim 0.001 Z_{\odot}$ (SOM).
As shown in Table \ref{tab:CII}, 
the derived excitation temperature for the levels $J= 3/2$ and $J=1/2$ of C\,{\sc
ii} for the damped Ly$\alpha$ absorber toward
Q0000$-$2619 at $z=3.0543$  is
between 19.6 and 21.6 K. We can use these values and
expression \ref{eq:final} to constrain the density and UV flux at the absorber.
However, in order to evaluate Eq.\,(\ref{eq:final}), we need to make an assumption
about the ionization degree of the gas in the absorber. 
Since the efficiency of the electron and hydrogen collisional  is very
different, the collisional excitation term depends critically on the
gas ionization degree.  To illustrate this we will
consider two limiting cases for the absorbing cloud: a quasi-neutral
gas, for which $n_e < 0.002~n_{\rm H}$, and
a 10\% ionized plasma.

\begin{figure}[htbp]
  \begin{center}
    \leavevmode \epsfig{file=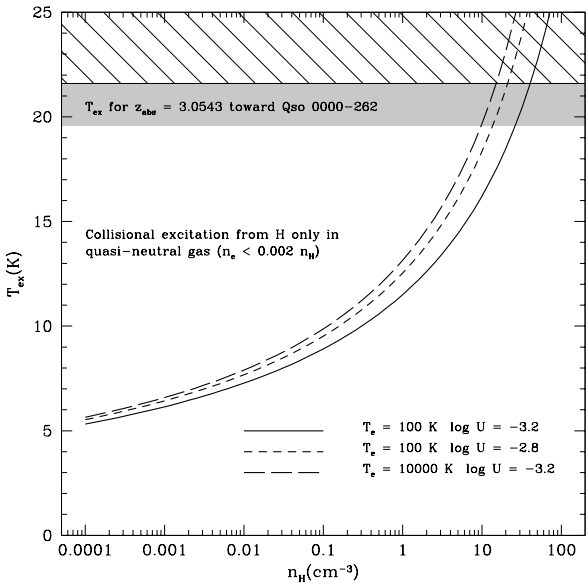, width=8.5cm, bbllx=10pt,
      bblly=150pt, bburx=600pt, bbury=700pt, clip=}
    \caption{The fine structure excitation temperature of C\,{\sc ii} as a
    function of the absorber density, under  the 
    assumption that the absorber is a quasi-neutral gas and collisional excitation is
    due to atomic hydrogen. The shaded area corresponds to the range
    of the derived excitation temperature for the damped Ly$\alpha$
    absorber at $z = 3.0543$  toward Q0000$-$2619. The continuous horizontal line gives
    the upper limit on the excitation temperature and implies that the
    density of this system must be lower than 40 cm$^{-3}$.}
    \label{fig:TexNeutral}
  \end{center}
\end{figure}

\begin{figure}[htbp]
  \begin{center}
    \leavevmode \epsfig{file=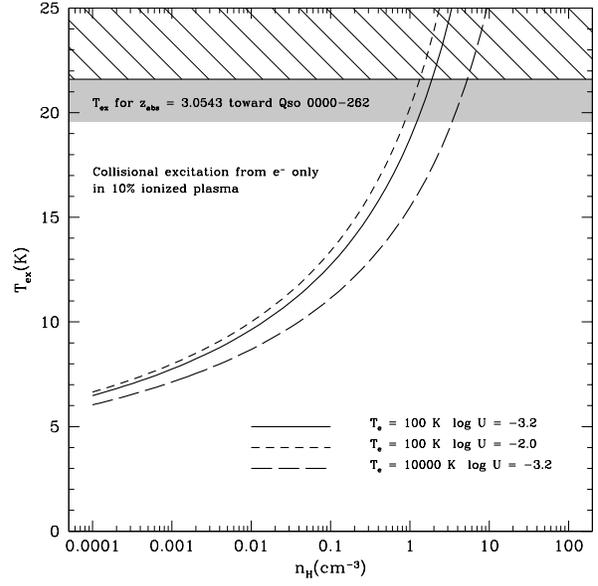, width=8.5cm, bbllx=10pt,
      bblly=150pt, bburx=600pt, bbury=700pt, clip=}
    \caption{Fine structure excitation temperature of C\,{\sc ii} as a
    function of the absorber density, under the
    assumption that 10\% of the gas is ionized and collisional excitation is
    due to electrons. The shaded area corresponds to the range
    of the derived excitation temperature for the damped Ly$\alpha$
    absorber at $z = 3.0543$  toward Q0000$-$2619.}
    \label{fig:TexIonized}
  \end{center}
\end{figure}

In Fig.\,\ref{fig:TexNeutral} the result of Eq.\,(\ref{eq:final})
is plotted, in the case of a quasi-neutral gas, where the collisional
excitation is due to collisions with atomic hydrogen.
The horizontal continuous line gives the upper limit on the excitation temperature.
The points where the curves intersect this line  correspond to the density values above 
which the collisional
excitation would be such  that the excitation temperature  would be
higher than the measured value. This is a forbidden
region. The density of the absorber must thus be lower than 40 cm$^{-3}$.
This is a strict upper limit. 
Higher values of the ionization parameters would move the intersection
point  to lower
density values. If the C\,{\sc ii}$^*$ absorbing line is a spurious
effect of the Lyman forest, then the C\,{\sc ii}$^*$  equivalent width must
be lower than the one we measured and the upper limit on the density
must be lower; increasing the cloud kinetic temperature also moves
the density-limit to lower values.
With $\log U = -2.8$ the upper limit on the absorber density is 20
cm$^{-3}$. 
The shaded area in the figure corresponds to the interval
of the derived excitation temperature for the absorber toward
Q0000$-$2619. The intersection of the long-dashed curve with the
lower bound of this area provides a lower limit of 10 ${\rm cm}^{-3}$
to the density of the system, 
if the gas is quasi-neutral ($n_{\rm e} < 0.002~{\rm n_H}$).

In Fig.\,\ref{fig:TexIonized}, C\,{\sc ii} fine structure excitation
temperature is plotted as a function of
the absorber density (like in Fig.\,\ref{fig:TexNeutral}) this time
for the case of a
10\% ionized plasma. Electron collisions dominate collisional
excitation in this case and because electron collisions are more
efficient than atomic collisions, lower gas densities are
required to achieve the same level of  C\,{\sc ii} fine structure
excitation. From Fig.\,\ref{fig:TexIonized} one derives that with ionization
parameter $\log U = -3.2$ 
%(thick--continuous lines) 
the range of possible gas
densities is between 1 and 5 ${\rm cm}^{-3}$. 
We note however that the exact value of the
ionization parameter has little effect on the position of the
intersect for the curves in Fig.\,\ref{fig:TexIonized}. For any $\log U \leq -2.0$
(thick--dotted line) 
we obtain that $n_{\rm H}$ must be greater than 0.7 ${\rm cm}^{-3}$, with the
same 10\% ionization and temperature range. An ionization parameter of
$-2.0$ at a density of 0.7 ${\rm cm}^{-3}$ correspond to a UV flux at
912 \AA\ of 20 times the Galactic UV field (if the same SED is assumed).  

These simple considerations allow us to constrain the absorber
density to the range of 0.7 $-$ 40 ${\rm cm}^{-3}$, for a
gas kinetic temperature between 100 and 10\,000 K and ionization degree
between 0 and 10\%. An ionization degree between 0 and 100\% would
imply possible absorber densities in the range 0.1 $-$ 40 ${\rm cm}^{-3}$.
An ionization parameter $\log U = -3.2$ and the density range  1.0 $-$
40 ${\rm cm}^{-3}$ imply a
UV field with H-ionizing photon flux ranging from 2 to 80 times the Galactic
H-ionizing flux. The upper value is a strict upper limit on the
intensity of the UV flux in the absorbing system.

\section{Discussion}
\label{sec:disc}

The observed excitation temperature (or upper limit) of C\,{\sc i} and
C\,{\sc ii} fine structure has been used to constrain the density
of the absorbing systems by several authors and the published values
are very similar to the upper limit derived here.
\cite*{Songaila94} observed absorption from  the first fine-structure level 
of C\,{\sc i} in a cloud belonging to a damped Ly$\alpha$ system
at a redshift of 1.776, towards the quasar Q1331$+$170. They measured
an excitation temperature of $7.4 \pm 0.8$ K, while the CMB temperature at
this redshift is predicted to be 7.58 K.
From their measure they derive a 1$\sigma$ upper limit for the cloud density of $n_{\rm H} = 7$ cm$^{-3}$
for a cloud kinetic temperature of 100 K and of $n_{\rm H} = 4$ cm$^{-3}$
for a cloud kinetic temperature of 1000 K, given that the CMB is at
the predicted temperature.
%If no assumption is made about the CMB temperature at that redshift the measured excitation
%temperature correspond to the upper limits for the cloud density of
%$n_{\rm H} = 26$ cm$^{-3}$, for a cloud kinetic temperature of 100 K and of
%$n_{\rm H} = 15$ cm$^{-3}$ for $T = 1000$ K.
\cite*{Ge97} detected absorption from the ground state and the excited
state of C\,{\sc i} and C\,{\sc ii} in the $z=1.9731$ damped Ly$\alpha$
system of Q0013$-$004. They measure an excitation temperature of $11.6
\pm 1.0$ K for C\,{\sc i} and of $16.1 \pm 1.4$ K for C\,{\sc ii}. They
use the C\,{\sc ii} excitation temperature to constrain the density of
the cloud to be $n_{\rm H} = 21.0 \pm 9.6$ cm$^{-3}$ if the cloud kinetic
temperature is of 100 K and $n_{\rm H} = 4.5 \pm 2.0$ cm$^{-3}$ for a kinetic
temperature of 1000 K.  With these densities and a photo-ionization
parameter $\log U = -3.5$, their estimate of the H-ionizing photon
flux ranges between 3.6 and 17 times the Galactic H-ionizing flux.
These values are then combined with
the measured  C\,{\sc i} excitation temperature to derive an upper
limit of $7.9 \pm 1.0$ K or $10.6 \pm 1.0$ K respectively on the
temperature of the CMB at this redshift, with the first value as their best
guess, based on photo-ionization modeling of the absorbing
cloud.

To constrain the density and UV field of the $z = 3.0543$ absorber of Q0000$-$2619
we have considered a gas kinetic temperature in the range 100$-$10\,000
K. There are not many observational constraints on the gas
temperature of damped Ly$\alpha$ systems. Cloud photo-ionization
models can reproduce the observational data with the gas temperature
ranging from 15 to 10\,000 K (\cite{Chaffee88}). However
observations of 21 cm absorption line from damped Ly$\alpha$ systems at
$z \sim 3$ obtained lower limits on the neutral hydrogen spin
temperature of the order of 1000 K (\cite{Carilli96};
\cite{Kanekar97}). This temperature must not be taken as a measure of
the gas mean temperature, but as an indication of the presence of warm
gas in the absorbing system. The typical spin
temperature measured in the Galactic clouds is $T_s \sim 100$ K
(\cite{Braun92}). Despite the Galactic cloud spin temperature not being
directly comparable with the
measure of the spin temperature in a damped Ly$\alpha$ system, a $T_s
\geq 1000$ K suggests that for a given total neutral hydrogen column
density, the  damped Ly$\alpha$ system contains a larger percentage of
warm phase gas ($T\sim 8000$ K) than is seen in typical Galactic
lines of sight (\cite{Carilli96}). 

The constraints on the density of the damped system can be used to
estimate its size.
In the simplistic hypothesis that the absorber is homogeneous a 
neutral hydrogen column density of $1.5 \pm 0.5 \times 10^{20}~{\rm
cm^{-2}}$ and a density range
of 0.7 $-$ 40 ${\rm cm}^{-3}$ imply that the size of the system along
the line of sight is between 1$-$100 pc. If the filling factor is significantly
lower than 1, this estimate is a lower limit.
It has been proposed that the DLAs with neutral
hydrogen column density $N$(H\,{\sc i})$ \geq  2 \times 10^{20}~{\rm
cm^{-2}}$, are large and massive galactic disks (e.g.,
\cite{Prochaska97}) with typical sizes of a few kpc.
The constraints we derived for this damped system at $z = 3.0543$ indicate
that the size of this absorber is of the order of the size
of giant hydrogen clouds in our galaxy or of a galactic disk seen
face-on. 
If this is not a galactic disk, at these redshift, 
systems of this scale could be protogalactic
clumps, that is the building blocks
of the various type of galaxies (ellipticals, spirals, etc.) that are
observed at present epoch (\cite{Khersonsky96}, \cite{Haehnelt98}).

\section{Conclusion}
\label{sec:conclusion}

In the damped Ly$\alpha$ system at $z=3.0543$ of Q0000$-$2619 studied
by SOM C\,{\sc ii} absorption was detected. In this paper we report
the detection at 3.5$\sigma$ of absorption from the excited fine structure level of C\,{\sc
ii} at $z=3.0543$. From the measure of the equivalent width of the two
lines we derived an upper limit of 21.6  K on the fine structure excitation
temperature. This value provides a strict upper limit on the temperature of
the CMB at $z=3.0543$, which at  this redshift is
predicted to be 11.05 K. We then used the derived relative
populations of the fine structure levels of C\,{\sc ii}
to set constraints on the absorber density and on the UV
field in the absorbing cloud. Assuming an ionization degree
ranging from 0 to 10\%, a gas kinetic temperature between 100
and 10\,000 K and a UV field with a Milky Way spectrum, 
the density of the absorber is constrained to be between 0.7 and 40
cm$^{-3}$ and the H-ionizing flux between $1\times 10^7$ and $8\times
10^8~{\rm cm^{-2}s^{-1}}$, that is between 1 and 80 times the intensity
of the Galactic UV field. The upper limits 
hold even 
if the detected absorption from C\,{\sc ii}$^*$ at $z=3.0543$
is seriously contaminated by Lyman forest absorption.
If the damped Ly$\alpha$ system is assumed to be homogeneous, a
density value between 0.7 and 40 cm$^{-3}$ constrains the size of
this absorber at $z \sim 3$ to be between 1 and 100 pc in the
direction of the line of sight.

\begin{acknowledgements}
 
We would like to thank S.~Savaglio for providing the NTT spectra of
Q0000$-$2619 and for her helpful comments to the manuscript, and 
P.~Jakobsen for many illuminating discussions.

\end{acknowledgements}


\begin{thebibliography}{}

\bibitem[\protect\astroncite{{Allen}}{1963}]{allen}
{Allen} C. 1963,
\newblock Astrophysical Quantities,
\newblock The Athlone Press (London)

\bibitem[\protect\astroncite{{Bahcall} \& {Wolf}}{1968}]{Bahcall68}
{Bahcall} J.~N., {Wolf} R.~A. 1968, ApJ 152, 701

\bibitem[\protect\astroncite{{Bennett} et~al.}{1992}]{Bennett92}
{Bennett} C.~L., {Smoot} G.~F., {Hinshaw} et~al. 1992, ApJ Lett. 396, L7

\bibitem[\protect\astroncite{{Black}}{1987}]{Black87}
{Black} J. 1987,
\newblock in Interstellar Processes,
\newblock ed. D.J. Hollenback \& H.A. Throsin, Jr (Dordrecht:Reidel), 931

\bibitem[\protect\astroncite{{Braun} \& {Walterbos}}{1992}]{Braun92}
{Braun} R., {Walterbos} R. A.~M. 1992, ApJ 386, 120

\bibitem[\protect\astroncite{{Carilli} et~al.}{1996}]{Carilli96}
{Carilli} C.~L., {Lane} W., {De Bruyn} A.~G. et~al. 1996, AJ 111, 1830

\bibitem[\protect\astroncite{{Chaffee} et~al.}{1988}]{Chaffee88}
{Chaffee}, F.~H. J., {Foltz} C.~B., {Black} J.~H. 1988, ApJ 335, 584

\bibitem[\protect\astroncite{{D'Odorico}}{1990}]{Dodorico90}
{D'Odorico} S. 1990, ESO The Messenger 61, 51

\bibitem[\protect\astroncite{{Ge} et~al.}{1997}]{Ge97}
{Ge} J., {Bechtold} J., {Black} J.~H. 1997, ApJ 474, 67

\bibitem[\protect\astroncite{{Gondhalekar} et~al.}{1980}]{Gondhalekar80}
{Gondhalekar} P.~M., {Phillips} A.~P., {Wilson} R. 1980, A\&A 85, 272

\bibitem[\protect\astroncite{{Haehnelt} et~al.}{1998}]{Haehnelt98}
{Haehnelt} M.~G., {Steinmetz} M., {Rauch} M. 1998, ApJ 495, 647

\bibitem[\protect\astroncite{{Hayes} \& {Nussbaumer}}{1984}]{Hayes84}
{Hayes} M.~A., {Nussbaumer} H. 1984, A\&A 134, 193

\bibitem[\protect\astroncite{{Kanekar} \& {Chengalur}}{1997}]{Kanekar97}
{Kanekar} N., {Chengalur} J.~N. 1997, MNRAS 292, 831

\bibitem[\protect\astroncite{{Keenan} et~al.}{1986}]{Keenan86}
{Keenan} F.~P., {Lennon} D.~J., {Johnson} C.~T. et~al. 1986, MNRAS 220, 571

\bibitem[\protect\astroncite{{Khersonsky} \& {Turnshek}}{1996}]{Khersonsky96}
{Khersonsky} V.~K., {Turnshek} D.~A. 1996, ApJ 471, 657

\bibitem[\protect\astroncite{{Kogut} et~al.}{1996}]{Kogut96}
{Kogut} A., {Banday} A.~J., {Bennett} et~al. 1996, ApJ 460, 1

\bibitem[\protect\astroncite{{Lu} et~al.}{1996}]{Lu96}
{Lu} L., {Sargent} W. L.~W., {Barlow} T.~A. et~al. 1996, ApJS 107, 475

\bibitem[\protect\astroncite{{Mathis} et~al.}{1983}]{Mathis83}
{Mathis} J.~S., {Mezger} P.~G., {Panagia} N. 1983, A\&A 128, 212

\bibitem[\protect\astroncite{{Morton}}{1991}]{Morton91}
{Morton} D.~C. 1991, ApJS 77, 119

\bibitem[\protect\astroncite{{Nussbaumer} \& {Storey}}{1981}]{Nussbaumer81}
{Nussbaumer} H., {Storey} P.~J. 1981, A\&A 96, 91

\bibitem[\protect\astroncite{Peebles}{1993}]{Peebles93}
Peebles P. 1993,
\newblock {\it Principle of physical cosmology},
\newblock Princeton University Press

\bibitem[\protect\astroncite{{Phillips}}{1994}]{Phillips94}
{Phillips} P.~R. 1994, MNRAS 269, 771

\bibitem[\protect\astroncite{{Prochaska} \& {Wolfe}}{1997}]{Prochaska97}
{Prochaska} J.~X., {Wolfe} A.~M. 1997, ApJ 487, 73

\bibitem[\protect\astroncite{{Savaglio} et~al.}{1994}]{Savaglio94}
{Savaglio} S., {D'Odorico} S., {Moller} P. 1994, A\&A 281, 331

\bibitem[\protect\astroncite{{Savaglio} et~al.}{1997}]{Savaglio97}
{Savaglio} S., {Cristiani} S., {D'Odorico} S. et~al. 1997, A\&A 318, 347

\bibitem[\protect\astroncite{{Songaila} et~al.}{1994}]{Songaila94}
{Songaila} A., {Cowie} L.~L., {Vogt} S. et~al. 1994, Nat 371, 43

\bibitem[\protect\astroncite{{Spitzer}}{1978}]{spitzer}
{Spitzer} L. 1978,
\newblock Physical Processes in the Interstellar Medium,
\newblock John Wiley \& Sons, Inc.

\bibitem[\protect\astroncite{{Sutherland} \& {Dopita}}{1993}]{Sutherland93}
{Sutherland} R.~S., {Dopita} M.~A. 1993, ApJS 88, 253

\end{thebibliography}
\end{document}